\documentclass[preprint]{aastex}
\newcommand{\etal}{et al.}
\begin{document}
\title{The Fourth Data Release of the Sloan Digital Sky Survey}
\author{
Jennifer K. Adelman-McCarthy\altaffilmark{\ref{Fermilab}},
Marcel A. Ag\"ueros\altaffilmark{\ref{Washington}},
Sahar S. Allam\altaffilmark{\ref{Fermilab},\ref{Wyoming}},
Kurt S. J. Anderson\altaffilmark{\ref{APO},\ref{NMSU}},
Scott F. Anderson\altaffilmark{\ref{Washington}},
James Annis\altaffilmark{\ref{Fermilab}},
Neta A. Bahcall\altaffilmark{\ref{Princeton}},
Ivan K. Baldry\altaffilmark{\ref{JHU}},
J. C. Barentine\altaffilmark{\ref{APO}},
Andreas Berlind\altaffilmark{\ref{NYU}},
Mariangela Bernardi\altaffilmark{\ref{Penn}}, 
Michael R. Blanton\altaffilmark{\ref{NYU}},
William N. Boroski\altaffilmark{\ref{Fermilab}},
Howard J. Brewington\altaffilmark{\ref{APO}},
Jarle Brinchmann\altaffilmark{\ref{Porto}},
J. Brinkmann\altaffilmark{\ref{APO}},
Robert J. Brunner\altaffilmark{\ref{Illinois}},
Tam\'as Budav\'ari\altaffilmark{\ref{JHU}},
Larry N. Carey\altaffilmark{\ref{Washington}},
Michael A. Carr\altaffilmark{\ref{Princeton}},
Francisco J. Castander\altaffilmark{\ref{Barcelona}},
A. J. Connolly\altaffilmark{\ref{Pitt}},
Istv\'an Csabai\altaffilmark{\ref{Eotvos},\ref{JHU}},
Paul C. Czarapata\altaffilmark{\ref{Fermilab}},
Julianne J. Dalcanton\altaffilmark{\ref{Washington}},
Mamoru Doi\altaffilmark{\ref{IoAUT}},
Feng Dong\altaffilmark{\ref{Princeton}},
Daniel J. Eisenstein\altaffilmark{\ref{Arizona}},
Michael L. Evans\altaffilmark{\ref{Washington}},
Xiaohui Fan\altaffilmark{\ref{Arizona}},
Douglas P. Finkbeiner\altaffilmark{\ref{Princeton}},
Scott D. Friedman\altaffilmark{\ref{STScI}},
Joshua A. Frieman\altaffilmark{\ref{Fermilab},\ref{Chicago},\ref{CfCP}},
Masataka Fukugita\altaffilmark{\ref{ICRRUT}},
Bruce Gillespie\altaffilmark{\ref{APO}},
Karl Glazebrook\altaffilmark{\ref{JHU}},
Jim Gray\altaffilmark{\ref{Microsoft}},
Eva K. Grebel\altaffilmark{\ref{Basel}},
James E. Gunn\altaffilmark{\ref{Princeton}},
Vijay K. Gurbani\altaffilmark{\ref{Fermilab},\ref{Lucent2}},
Ernst de Haas\altaffilmark{\ref{Princeton}},
Patrick B. Hall\altaffilmark{\ref{York}},
Frederick H. Harris\altaffilmark{\ref{NOFS}},
Michael Harvanek\altaffilmark{\ref{APO}},
Suzanne L. Hawley\altaffilmark{\ref{Washington}},
Jeffrey Hayes\altaffilmark{\ref{Catholic}},
John S. Hendry\altaffilmark{\ref{Fermilab}},
Gregory S. Hennessy\altaffilmark{\ref{USNO}},
Robert B. Hindsley\altaffilmark{\ref{NRL}},
Christopher M. Hirata\altaffilmark{\ref{Princetonphys}},
Craig J. Hogan\altaffilmark{\ref{Washington}},
David W. Hogg\altaffilmark{\ref{NYU}},
Donald J. Holmgren\altaffilmark{\ref{Fermilab}},
Jon A. Holtzman\altaffilmark{\ref{NMSU}},
Shin-ichi Ichikawa\altaffilmark{\ref{NAOJ}},
\v{Z}eljko Ivezi\'{c}\altaffilmark{\ref{Washington}},
Sebastian Jester\altaffilmark{\ref{Fermilab}},
David E. Johnston\altaffilmark{\ref{Princeton}},
Anders M. Jorgensen\altaffilmark{\ref{LANL}},
Mario Juri\'{c}\altaffilmark{\ref{Princeton}},
Stephen M. Kent\altaffilmark{\ref{Fermilab}},
S. J. Kleinman\altaffilmark{\ref{APO}},
G. R. Knapp\altaffilmark{\ref{Princeton}},
Alexei Yu. Kniazev\altaffilmark{\ref{MPIA}},
Richard G. Kron\altaffilmark{\ref{Chicago},\ref{Fermilab}},
Jurek Krzesinski\altaffilmark{\ref{APO},\ref{MSO}},
Nikolay Kuropatkin\altaffilmark{\ref{Fermilab}},
Donald Q. Lamb\altaffilmark{\ref{Chicago},\ref{EFI}},
Hubert Lampeitl\altaffilmark{\ref{Fermilab}},
Brian C. Lee\altaffilmark{\ref{LBL}},
R. French Leger\altaffilmark{\ref{Fermilab}},
Huan Lin\altaffilmark{\ref{Fermilab}},
Daniel C. Long\altaffilmark{\ref{APO}},
Jon Loveday\altaffilmark{\ref{Sussex}},
Robert H. Lupton\altaffilmark{\ref{Princeton}},
Bruce Margon\altaffilmark{\ref{STScI}},
David Mart\'{\i}nez-Delgado\altaffilmark{\ref{Andalucia}},
Rachel Mandelbaum\altaffilmark{\ref{Princetonphys}},
Takahiko Matsubara\altaffilmark{\ref{Nagoya}},
Peregrine M. McGehee\altaffilmark{\ref{LANL2}},
Timothy A. McKay\altaffilmark{\ref{Michigan}},
Avery Meiksin\altaffilmark{\ref{Edinburgh}},
Jeffrey A. Munn\altaffilmark{\ref{NOFS}},
Reiko Nakajima\altaffilmark{\ref{Penn}},
Thomas Nash\altaffilmark{\ref{Fermilab}},
Eric H. Neilsen, Jr.\altaffilmark{\ref{Fermilab}},
Heidi Jo Newberg\altaffilmark{\ref{RPI}},
Peter R. Newman\altaffilmark{\ref{APO}},
Robert C. Nichol\altaffilmark{\ref{Portsmouth}},
Tom Nicinski\altaffilmark{\ref{Fermilab},\ref{CMCElectronics}},
Maria Nieto-Santisteban\altaffilmark{\ref{JHU}},
Atsuko Nitta\altaffilmark{\ref{APO}},
William O'Mullane\altaffilmark{\ref{JHU}},
Sadanori Okamura\altaffilmark{\ref{DoAUT}},
Russell Owen\altaffilmark{\ref{Washington}},
Nikhil Padmanabhan\altaffilmark{\ref{Princetonphys}},
George Pauls\altaffilmark{\ref{Princeton}},
John Peoples Jr.\altaffilmark{\ref{Fermilab}},
Jeffrey R. Pier\altaffilmark{\ref{NOFS}},
Adrian C. Pope\altaffilmark{\ref{JHU}},
Dimitri Pourbaix\altaffilmark{\ref{Princeton},\ref{Bruxelles}},
Thomas R. Quinn\altaffilmark{\ref{Washington}},
Gordon T. Richards\altaffilmark{\ref{Princeton}},
Michael W. Richmond\altaffilmark{\ref{RIT}},
Constance M. Rockosi\altaffilmark{\ref{Lick}},
David J. Schlegel\altaffilmark{\ref{LBL}},
Donald P. Schneider\altaffilmark{\ref{PSU}},
Joshua Schroeder\altaffilmark{\ref{Princeton},\ref{Colorado}},
Ryan Scranton\altaffilmark{\ref{Pitt}},
Uro\v{s} Seljak\altaffilmark{\ref{Princetonphys},\ref{Trieste}},
Erin Sheldon\altaffilmark{\ref{Chicago},\ref{CfCP}},
Kazu Shimasaku\altaffilmark{\ref{DoAUT}},
J. Allyn Smith\altaffilmark{\ref{Wyoming},\ref{LANL}},
Vernesa Smol\v{c}i\'{c}\altaffilmark{\ref{Zagreb}},
Stephanie A. Snedden\altaffilmark{\ref{APO}},
Chris Stoughton\altaffilmark{\ref{Fermilab}},
Michael A. Strauss\altaffilmark{\ref{Princeton}},
Mark SubbaRao\altaffilmark{\ref{Chicago},\ref{Adler}},
Alexander S. Szalay\altaffilmark{\ref{JHU}},
Istv\'an Szapudi\altaffilmark{\ref{Hawaii}},
Paula Szkody\altaffilmark{\ref{Washington}},
Max Tegmark\altaffilmark{\ref{MIT}},
Aniruddha R. Thakar\altaffilmark{\ref{JHU}},
Douglas L. Tucker\altaffilmark{\ref{Fermilab}},
Alan Uomoto\altaffilmark{\ref{JHU},\ref{CarnegieObs}},
Daniel E. Vanden Berk\altaffilmark{\ref{PSU}},
Jan Vandenberg\altaffilmark{\ref{JHU}},
Michael S. Vogeley\altaffilmark{\ref{Drexel}},
Wolfgang Voges\altaffilmark{\ref{MPIEP}},
Nicole P. Vogt\altaffilmark{\ref{NMSU}},
Lucianne M. Walkowicz\altaffilmark{\ref{Washington}},
David H. Weinberg\altaffilmark{\ref{OSU}},
Andrew A. West\altaffilmark{\ref{Washington}},
Simon D.M. White\altaffilmark{\ref{MPA}},
Yongzhong Xu\altaffilmark{\ref{LANLtheory}},
Brian Yanny\altaffilmark{\ref{Fermilab}},
D. R. Yocum\altaffilmark{\ref{Fermilab}},
Donald G. York\altaffilmark{\ref{Chicago},\ref{EFI}},
Idit Zehavi\altaffilmark{\ref{Arizona}},
Stefano Zibetti\altaffilmark{\ref{MPIEP}},
Daniel B. Zucker\altaffilmark{\ref{MPIA}}
}

\altaffiltext{1}{
Fermi National Accelerator Laboratory, P.O. Box 500, Batavia, IL 60510.
\label{Fermilab}}

\altaffiltext{2}{
Department of Astronomy, University of Washington, Box 351580, Seattle, WA
98195.
\label{Washington}}

\altaffiltext{3}{
Department of Physics and Astronomy, University of Wyoming, Laramie, WY 82071.
\label{Wyoming}}

\altaffiltext{4}{
Apache Point Observatory, P.O. Box 59, Sunspot, NM 88349.
\label{APO}}

\altaffiltext{5}{
Department of Astronomy, MSC 4500, New Mexico State University,
P.O. Box 30001, Las Cruces, NM 88003.
\label{NMSU}}

\altaffiltext{6}{
Department of Astrophysical Sciences, Princeton University, Princeton, NJ
08544.
\label{Princeton}}

\altaffiltext{7}{
Center for Astrophysical Sciences, Department of Physics and Astronomy, Johns
Hopkins University, 3400 North Charles Street, Baltimore, MD 21218. 
\label{JHU}}

\altaffiltext{8}{
Center for Cosmology and Particle Physics,
Department of Physics,
New York University,
4 Washington Place,
New York, NY 10003.
\label{NYU}}

\altaffiltext{9}{
Department of Physics and Astronomy, University of Pennsylvania,
Philadelphia, PA 19104. 
\label{Penn}}

\altaffiltext{10}{
Centro de Astrof{\'\i}sica da Universidade do Porto, Rua 
das Estrelas - 4150-762 Porto, Portugal.
\label{Porto}}

\altaffiltext{11}{
Department of Astronomy
University of Illinois
1002 West Green Street, Urbana, IL 61801.
\label{Illinois}}

\altaffiltext{12}{Institut d'Estudis Espacials de Catalunya/CSIC, Gran Capit\'a 2-4,
E-08034 Barcelona, Spain.
\label{Barcelona}}

\altaffiltext{13}{
Department of Physics and Astronomy, University of Pittsburgh, 3941 O'Hara
Street, Pittsburgh, PA 15260.
\label{Pitt}}

\altaffiltext{14}{
Department of Physics of Complex Systems, E\"{o}tv\"{o}s Lor\'and University, Pf.\ 32,
H-1518 Budapest, Hungary.
\label{Eotvos}}

\altaffiltext{15}{Institute of Astronomy and Research Center for the
  Early Universe, School
of Science, University of Tokyo,
 2-21-1 Osawa, Mitaka, Tokyo 181-0015, Japan.
\label{IoAUT}}

\altaffiltext{16}{
Steward Observatory, 933 North Cherry Avenue, Tucson, AZ 85721.
\label{Arizona}}

\altaffiltext{17}{
Space Telescope Science Institute, 3700 San Martin Drive, Baltimore, MD
21218.
\label{STScI}}

\altaffiltext{18}{
Department of Astronomy and Astrophysics, University of Chicago, 5640 South
Ellis Avenue, Chicago, IL 60637.
\label{Chicago}}

\altaffiltext{19}{
Kavli Institute for Cosmological Physics, The University of Chicago,
5640 South Ellis Avenue, Chicago, IL 60637.
\label{CfCP}}

\altaffiltext{20}{Institute for Cosmic Ray Research, University of Tokyo, 5-1-5 Kashiwa,
 Kashiwa City, Chiba 277-8582, Japan.
\label{ICRRUT}}

\altaffiltext{21}{
Microsoft Research, 455 Market Street, Suite 1690, San Francisco, CA 94105.
\label{Microsoft}}

\altaffiltext{22}{
Astronomical Institute of the University of Basel, 
Department of Physics and Astronomy, Venusstrasse 7, CH-4102 Basel,
Switzerland
\label{Basel}}

\altaffiltext{23}{
Lucent Technologies, 2000 Lucent Lane, Naperville, IL 60566.
\label{Lucent2}}

\altaffiltext{24}{
Dept. of Physics \& Astronomy,
York University,
4700 Keele St.,
Toronto, ON, M3J 1P3,
Canada
\label{York}}

\altaffiltext{25}{
US Naval Observatory, Flagstaff Station, 10391 W. Naval Observatory Road, Flagstaff, AZ
86001-8521.
\label{NOFS}}

\altaffiltext{26}{
Institute for Astronomy and Computational Sciences
     Physics Department
     Catholic University of America
     Washington DC 20064
\label{Catholic}}

\altaffiltext{27}{
US Naval Observatory, 3540 Massachusetts Avenue NW, Washington, DC 20392.
\label{USNO}}

\altaffiltext{28}{
Code 7215, Remote Sensing Division
Naval Research Laboratory
4555 Overlook Avenue SW
Washington, DC 20392.
\label{NRL}}

\altaffiltext{29}{
Joseph Henry Laboratories, Princeton University, Princeton, NJ
08544.
\label{Princetonphys}}

\altaffiltext{30}{National Astronomical Observatory, 2-21-1 Osawa, Mitaka, Tokyo 181-8588,
Japan.
\label{NAOJ}}

\altaffiltext{31}{
ISR-4, MS D448, Los Alamos National Laboratory, P.O.Box 1663, Los Alamos, NM 87545.
\label{LANL}}

\altaffiltext{32}{
Max-Planck-Institut f\"ur Astronomie, K\"onigstuhl 17, D-69117 Heidelberg,
Germany.
\label{MPIA}}

\altaffiltext{33}{
Obserwatorium Astronomiczne na Suhorze, Akademia Pedogogiczna w
Krakowie, ulica Podchor\c{a}\.{z}ych 2,
PL-30-084 Krac\'ow, Poland.
\label{MSO}}

\altaffiltext{34}{
Enrico Fermi Institute, University of Chicago, 5640 South Ellis Avenue,
Chicago, IL 60637.
\label{EFI}}

\altaffiltext{35}{
Lawrence Berkeley National Laboratory, One Cyclotron Road,
Berkeley CA 94720-8160.
\label{LBL}}

\altaffiltext{36}{
Astronomy Centre, University of Sussex, Falmer, Brighton BN1 9QJ, UK. 
\label{Sussex}}

\altaffiltext{37}{
Instituto de Astrofisica de Andalucia 
(CSIC), Camino Bajo de Huetor, 24 18008 Granada, Spain.
\label{Andalucia}}

\altaffiltext{38}{
Department of Physics and Astrophysics,
 Nagoya University,
 Chikusa, Nagoya 464-8602,
 Japan.
\label{Nagoya}}

\altaffiltext{39}{
LANSCE-8, MS H820, Los Alamos National Laboratory, P.O.Box 1663, Los Alamos, NM 87545.
\label{LANL2}}

\altaffiltext{40}{
Department of Physics, University of Michigan, 500 East University Avenue, Ann
Arbor, MI 48109.
\label{Michigan}}

\altaffiltext{41}{
Institute for Astronomy,
Royal Observatory,
University of Edinburgh,
Blackford Hill,
Edinburgh EH9 3HJ,
UK.
\label{Edinburgh}}

\altaffiltext{42}{
Department of Physics, Applied Physics, and Astronomy, Rensselaer
Polytechnic Institute, 110 Eighth Street, Troy, NY 12180. 
\label{RPI}}

\altaffiltext{43}{
Institute of Cosmology and Gravitation (ICG),
Mercantile House, Hampshire Terrace,
Univ. of Portsmouth, Portsmouth, PO1 2EG, UK.
\label{Portsmouth}}

\altaffiltext{44}{
    CMC Electronics Aurora,
 84 N. Dugan Rd.
    Sugar Grove, IL 60554.
\label{CMCElectronics}}

\altaffiltext{45}{Department of Astronomy and Research Center for the Early Universe, 
University of Tokyo,
 7-3-1 Hongo, Bunkyo, Tokyo 113-0033, Japan.
\label{DoAUT}}

\altaffiltext{46}{
FNRS
Institut  d'Astronomie et d'Astrophysique,
 Universit\'e Libre de Bruxelles, CP. 226, Boulevard du Triomphe, B-1050
 Bruxelles, Belgium.
\label{Bruxelles}}

\altaffiltext{47}{
Department of Physics, Rochester Institute of Technology, 84 Lomb Memorial
Drive, Rochester, NY 14623-5603.
\label{RIT}}

\altaffiltext{48}{
UCO/Lick Observatory, University of California, Santa Cruz, CA 95064.
\label{Lick}}

\altaffiltext{49}{
Department of Astronomy and Astrophysics, 525 Davey Laboratory, Pennsylvania State
University, University Park, PA 16802.
\label{PSU}}

\altaffiltext{50}{
Center for Astrophysics and Space Astronomy, University of Colorado,
Boulder, CO 80309.
\label{Colorado}}

\altaffiltext{51}{
International Centre for Theoretical Physics
Strada Costiera 11
I-34014 Trieste, Italy.
\label{Trieste}}

\altaffiltext{52}{University of Zagreb, 
Department of Physics, Bijeni\v{c}ka cesta 32, 
10000 Zagreb, Croatia.
\label{Zagreb}}

\altaffiltext{53}{
Adler Planetarium and Astronomy Museum,
1300 Lake Shore Drive,
Chicago, IL 60605.
\label{Adler}}

\altaffiltext{54}{
Institute for Astronomy, 2680 Woodlawn Road, Honolulu, HI 96822.
\label{Hawaii}}

\altaffiltext{55}{
Dept. of Physics, Massachusetts Institute of Technology, Cambridge,  
MA 02139.
\label{MIT}}

\altaffiltext{56}{
Observatories of the Carnegie Institution of Washington, 
813 Santa Barbara Street, 
Pasadena, CA  91101.
\label{CarnegieObs}}

\altaffiltext{57}{
Department of Physics, Drexel University, 3141 Chestnut Street, Philadelphia, PA 19104.
\label{Drexel}}

\altaffiltext{58}{
Max-Planck-Institut f\"ur extraterrestrische Physik, 
Giessenbachstrasse 1, D-85741 Garching, Germany.
\label{MPIEP}}

\altaffiltext{59}{
Department of Astronomy, Ohio State University, 140 West 8th Avenue, Columbus, OH 43210.
\label{OSU}}

\altaffiltext{60}{
Max Planck Institut f\"ur Astrophysik, Postfach 1, 
D-85748 Garching, Germany.
\label{MPA}}

\altaffiltext{61}{
Theoretical Division, MS B285, Los Alamos National Laboratory, Los Alamos, NM 87545.
\label{LANLtheory}}

\shorttitle{SDSS DR4}
\shortauthors{Adelman-McCarthy \etal}

\begin{abstract}
This paper describes the fourth data release of the Sloan Digital Sky
Survey (SDSS), including all survey-quality data taken through June
2004.  The data release includes five-band photometric data for 180 million
objects selected over 6670 deg$^2$, and 673,280 spectra of
galaxies, quasars, and stars selected from 4783 deg$^2$ of
that imaging data using the standard SDSS target selection algorithms.
These numbers represent a roughly 25\% increment over those of the
Third Data Release.  The Fourth Data Release also includes an
additional 131,840 spectra of objects selected using a variety of
alternative algorithms, to address scientific issues ranging from the
kinematics of stars in the Milky Way thick disk to populations of
faint galaxies and quasars.
\end{abstract}
\section{Introduction}
\label{sec:introduction}

The Sloan Digital Sky Survey (SDSS) is an imaging and spectroscopic
survey of the sky (York \etal\ 2000) using a dedicated wide-field 2.5m
telescope (Gunn \etal\ 2005) at Apache Point Observatory, New Mexico.
Imaging is carried out in drift-scan mode using a 142 mega-pixel
camera (Gunn \etal\ 1998) which gathers data in five broad bands,
$u\,g\,r\,i\,z$, 
spanning the range from 3000 to 10,000 \AA\ (Fukugita \etal\ 1996), on
moonless photometric (Hogg \etal\ 2001) nights of good seeing. 
The images are processed using specialized software (Lupton \etal\
2001; Lupton 2005; Stoughton \etal\ 2002), and are astrometrically (Pier
\etal\ 2003) and photometrically (Tucker \etal\ 2005) calibrated using
observations of a set of primary standard stars (Smith \etal\ 2002)
on a neighboring 20-inch telescope.  The photometric calibration is
accurate to roughly 2\% rms in the $g, r$ and $i$ bands, and 3\% in $u$ and
$z$, as determined by the constancy of stellar population colors
(Ivezi\'c \etal\ 2004b; Blanton \etal\ 2005a), while the astrometric
calibration precision is better than 0.1 arcsec rms per coordinate
(Pier \etal\ 2003).  The median seeing of the imaging data is 1.4
arcsec in the $r$ band, and the 95\% completeness limit in
the $r$ band is 22.2.  All magnitudes are roughly on an AB system
(Abazajian \etal\ 2004), and use the asinh scale described by Lupton,
Gunn, \& Szalay (1999).  

  Objects are selected from the imaging data for spectroscopy using a
  variety of algorithms, including a complete sample of galaxies with
  reddening-corrected (Schlegel, Finkbeiner, \& Davis 1998; SFD)
  Petrosian (1976) 
  $r$ magnitudes brighter than 17.77 (Strauss \etal\ 2002), a deeper
  sample of color- and magnitude-selected galaxies targeting luminous
  red galaxies (LRGs) from redshift 0.15 to beyond 0.5 (Eisenstein \etal\
  2001), a color-selected sample of quasars with $0 < z < 5.5$
  (Richards \etal\ 2002), optical counterparts to ROSAT X-ray
  sources (Anderson \etal\ 2003), and a variety of stellar and
  calibrating objects (Stoughton \etal\ 2002).  These targets are
  arranged on a series of 
  tiles of radius $1.49^\circ$ (Blanton \etal\ 2003), each containing
  640 targets and calibration objects, and holes are
  drilled in the corresponding positions in aluminum plates.  Optical
  fibers with diameter $3''$ at the focal plane are plugged into these
  holes, and feed the light of the targeted objects to a pair of
  double spectrographs.  
  The resulting spectra cover the wavelength range $3800-9200$ \AA\ with
  a resolution of $\lambda/\Delta \lambda \approx 2000$.  The spectra
  are flux-calibrated using F subdwarfs, with a broad-band uncertainty
  of 5\% (Abazajian \etal\ 2004); the wavelength calibration
  uncertainty is roughly 0.05 \AA. 
  More than 98\% of all spectra are of high enough quality to yield an
  unambiguous classification and redshift.  

  The SDSS began formal operations in April 2000.  Commissioning data
  taken before that time were released in an Early Data Release
  (EDR; Stoughton \etal\ 2002), and subsequent data have been released in
  roughly yearly intervals (the first, second, and third data
  releases, DR1, DR2, and DR3; Abazajian \etal\ 2003, 2004, 2005).
  This paper describes the fourth data 
  release, hereafter DR4, consisting of survey data meeting our
  quality specifications taken through June 2004.  Data releases are
  cumulative; DR4 includes all data released in previous data
  releases.  There have been no substantive changes to the imaging or
  spectroscopic software since DR2. 

  Finkbeiner \etal\ (2004) describe the release of SDSS imaging data
  taken outside the formal SDSS footprint (as described by York \etal\
  2000); most of these data are at low Galactic latitudes.  In addition,
  there has been a number of value-added catalogs of SDSS data,
  including galaxies (Blanton \etal\ 2005a), quasars (Schneider \etal\
  2005), clusters of galaxies (Miller \etal\ 2005), compact groups
  (Lee \etal\ 2004), merging galaxies (Allam \etal\ 2004), white dwarf stars
  (Kleinman \etal\ 2004), and asteroids (Ivezi\'c
  \etal\ 2004a). 

  Section~\ref{sec:DR4} of this paper describes the fourth data release
  itself.  This release includes spectra from 209 plates principally
  in the Southern Galactic Cap which explore a variety of extensions
  of the main spectroscopic targeting algorithms; these are described
  in \S~\ref{sec:southern}.  Additional new features of DR4 can be
  found in \S~\ref{sec:new}.  Section~\ref{sec:caveats} describes a
  recently discovered subtle effect in the SDSS data, whereby
  mis-estimated sky levels near bright galaxies systematically bias
  the counts of neighboring faint galaxies. We conclude in
  \S~\ref{sec:conclusions}.  

\section{What is included in DR4}
\label{sec:DR4}
\begin{figure}
\plotone{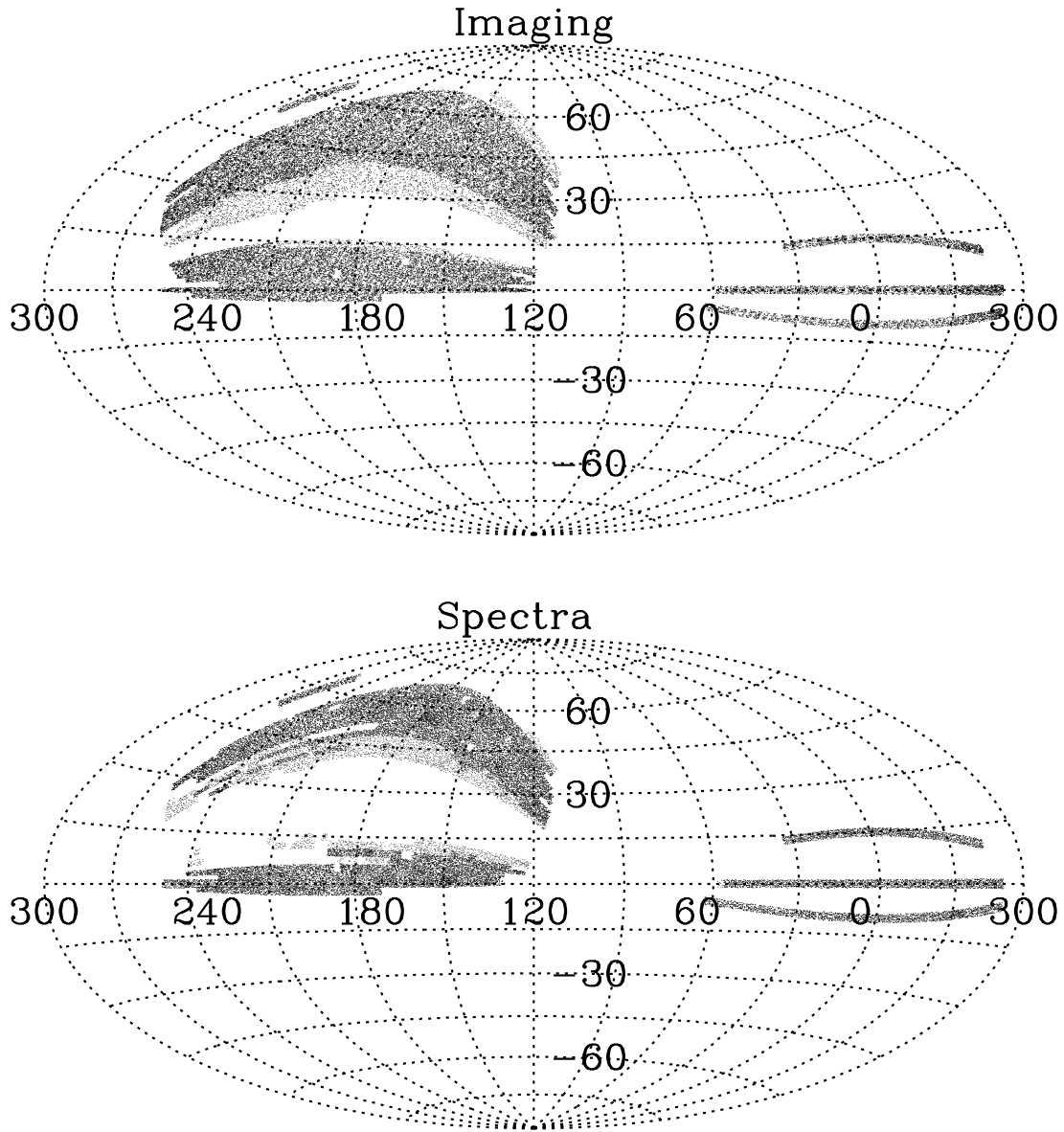}
\caption{
The distribution on the sky of SDSS imaging (upper panel) and
spectroscopy (lower panel) included in DR4, shown in J2000 equatorial
coordinates.  These cover 6670 and 4783 deg$^2$, respectively.  The
regions of sky that are new to DR4 are shaded more lightly.  
\label{fig:skydist}}\end{figure}

As is described by York \etal\ (2000), the SDSS imaging data are taken
along a series of stripes, great circles on the sky which aim to fill
a contiguous area in the Northern Galactic Cap, and three
non-contiguous stripes in the Southern Galactic Cap.  The top panel of
Figure~\ref{fig:skydist} shows the region of sky included in DR4.  As
this figure shows, the Northern Galactic Cap is currently covered by
two contiguous regions, one centered roughly on the Celestial Equator,
and the other at around $\delta = +40^\circ$.  New data
taken since DR3 (indicated in grey in the figure) have reduced the
size of the gap between these regions; these new data lie on the 
northern edge of the equatorial patch or on the southern edge of the
Northern patch.  The images cover 6670 deg$^2$.  The great circle
stripes overlap at the poles of the survey; roughly 25\% 
of the region is covered more than once; the repeat data in the
Northern Galactic Cap are
available in DR4 as well. 

These data are processed by the same version of the imaging pipelines
used in DR2 and DR3.  There have been concerns in previous data
releases that the SDSS imaging flat-fields change with time,
especially in $u$.  We have confirmed that the flat fields used for
the data taken since DR3 are correct to 2\% or better. 

The lower panel of Figure~\ref{fig:skydist} shows the sky coverage of the
spectroscopic data.  As the spectroscopy necessarily lags the imaging,
it covers less area, a total of 4783
deg$^2$.   Calculation of this area is more accurate than in previous
data releases (indeed Abazajian \etal\ (2003) erroneously indicated the
  spectroscopic area of DR3 to be 4188 deg$^2$; the correct value for
  DR3 is 3732 deg$^2$). The SDSS Catalog Archive Server (CAS) database
  includes a package that computes the spherical polygons corresponding
to all geometric objects in the survey, defined by the boundaries of
the imaging runs and the spectroscopic plates (Gray \etal\ 2004; see
also the discussion by Blanton \etal\ 2005a)\footnote{This package is
  not yet publically available with DR4, but will be included in a
  future data release.}.  Given plate overlaps,
one can divide up the spectroscopically tiled sky into non-overlapping
``sectors'', each one of which defines a region uniformly covered by
the same tiles.  Summing the area of those sectors which cover an
observed spectroscopic plate gives the total area of the
spectroscopic sample. 

  The spectroscopic data include 673,280 spectra,
arrayed on 1052 plates of 640 fibers each.  Thirty-two fibers per plate are
devoted to measurements of sky.  There are approximately 480,000 galaxies,
64,000 quasars, and 89,000 stars among the spectra; only
1\% of the spectra are unclassified.  About 2\%
of the objects are repeated on adjacent plates, as a check of the
reproducibility of the spectroscopy.  

  A number of spectroscopic plates have been observed more than once;
  this data release includes those duplicate observations which satisfy our
  signal-to-noise ratio (S/N) criteria.  These criteria
  are a mean S/N per pixel of 4 for objects with $g_{\rm fiber} = 20.2$
in the blue cameras of the spectrographs, and similarly for
  $i_{\rm fiber} = 19.9$ in the red cameras.  In particular,
  there are 52 plates with more than one observation which are
  released.  The majority of these plates have only one duplicate
  observation, but three plates (406, 419, and 483) have two
  duplicates, and two plates (406 and 412) have four. Note
  that in most cases, the plates were 
  replugged before being re-observed, so the
  correspondence between fiber number and object will be different on
  each observation.   These duplicate plates are useful for monitoring
  the robustness of 
  specific spectroscopic features, especially in low S/N spectra
  (e.g., Hall \etal\ 2004), increasing the S/N
  of spectra by co-adding, and checking for variability (Wilhite \etal\ 2005,
  although note that this paper used many spectroscopic observations that fall
  below our S/N cut, and are therefore not included in the release).  

  In the Fall months, when the Southern Galactic Cap is visible in the
  Northern Hemisphere, the SDSS imaging has been confined to a stripe
  along the Celestial Equator, plus two ``outrigger'' stripes,
  centered roughly at $\delta = +15^\circ$ and $\delta = -10^\circ$,
  respectively (these are visible on the right-hand-side of the 
  panels of Figure~\ref{fig:skydist}).  We have performed multiple
  imaging passes of the 
  Equatorial stripe, data which are being used for a deep co-addition,
  as well as a variety of variability studies (e.g., Ivezi\'c \etal\
  2003).  These multiple imaging scans are planned for inclusion in a
  future release. 

\begin{figure}
\plotone{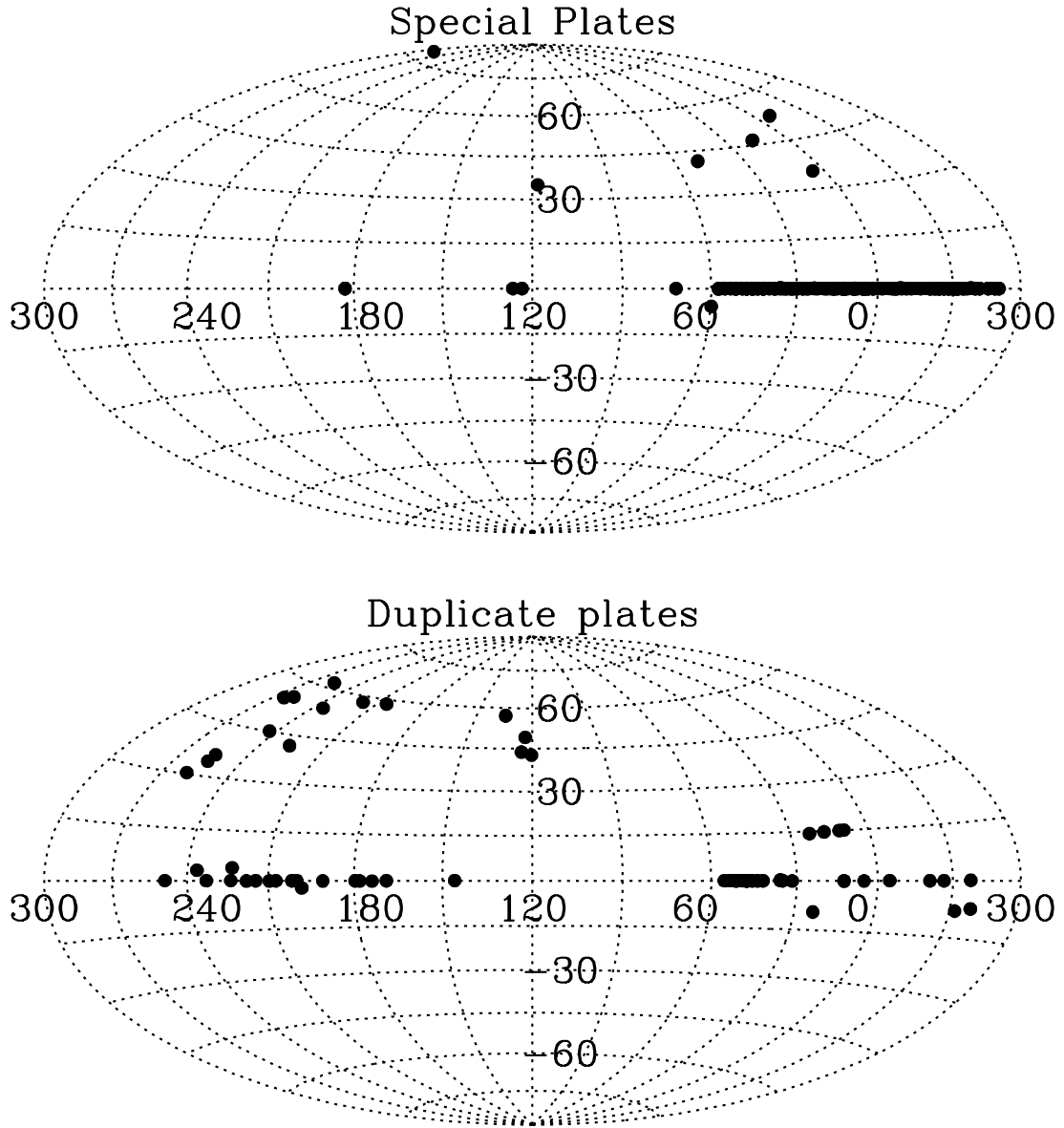}
\caption{
The distribution on the sky of special plates (upper panel) and
and duplicate plates (lower panel) included in DR4, shown in J2000 equatorial
coordinates.  The majority of the special plates are on the Celestial
Equator, in the Southern Equatorial Cap, but there are several
exceptions, as this figure makes clear. 
\label{fig:skydist_special}}
\end{figure}

  Given the relatively small footprint of the imaging in the Southern
  Galactic Cap, we finished the spectroscopy of targets selected by
  our normal algorithms quite early in the survey; most of these data
  were included already in DR1.  We carry out imaging only
  under pristine conditions: the moon is below the horizon, the sky is
  cloudless, 
  and the seeing is good.  To make optimal use of the remaining time,
  we undertook a series of 
  spectroscopic observing programs, based mostly on the imaging data
  of the Equatorial Stripe in the Southern Galactic Cap, designed to
  go beyond the science goals of the main survey.  DR4 includes 206
  plates, totaling 131,840 spectra, from these programs,
  carried out in the Fall months of 2001, 2002, and 2003.  These
  programs are described in the next section.  In addition, DR4
  includes nine plates which are repeat observations of these
  ``southern'' plates.  The sky distribution of both the special and
  repeat plates are given in Figure~\ref{fig:skydist_special}. 

  As described by Stoughton \etal\ (2002), these data are available
both as flat files (the Data Archive Server), and via a flexible
web interface to the SDSS data base (the
CAS)\footnote{http://www.sdss.org/dr4}.   

\section{The Special Southern Spectroscopic Plates}
\label{sec:southern}

Unlike the spectroscopic plates for the main survey, the special
southern plates did not aim to produce a complete, uniform survey of
targets over the entire survey area.  Indeed, many of the programs carried out
with these plates were quite exploratory in nature, designed to probe
the limits of the main survey target algorithms, push deeper in
various categories of targets, or select classes of objects not fully
explored by the main algorithms (especially stars).  These selection 
algorithms changed several times in the course of the 2001-2003
time period, which makes keeping track of the target selection flags
somewhat complicated.  Table~\ref{table:Southern_plates} shows which
algorithms (as described below) are used on which spectroscopic
plates.  This table makes reference to a {\it chunk}, i.e., a region
of imaging data on which target selection was run and spectroscopic
tiles defined (Blanton \etal\ 2003).  The different
chunks are numbered, but are also given names (e.g., {\tt lowz97}),
which can be useful in queries of the CAS.  While some chunks include only a
few plates, others include several tens of plates.  Among the most important of
these are the various merged programs.  Chunk 22, the first of the
major Southern programs, used simple extensions of the main survey
targeting algorithms (although the plates included some objects
targeted with the original algorithms as well).  Chunks 48 and 73
combined a variety of 
algorithms to target a range of different targets.  Chunks 45, 52,
62, 74 and 97 used double-length exposures to target faint galaxies
and stars.  

  As described by Stoughton \etal\ (2002), the
survey marks each object targeted for spectroscopy with flag bits
encoded in the primTarget and secTarget quantities.  The relevant flag
hex bits for the Southern plates are given in 
Table~\ref{table:southern_flags}.  With a few exceptions, these
special plates are flagged with the highest bit in secTarget.  The
primTarget flags unfortunately can take on different meanings
depending on chunk.  Thus using these target flags requires
taking the chunk number or plate numbers into account.
Table~\ref{table:southern_flags} also indicates the total number of
objects with each of the target flags. 

All of the programs described below put a bright limit of 15 in $g$,
$r$, and/or $i$ fiber magnitudes (i.e., aperture magnitudes of
$3^{\prime\prime}$ diameter, corresponding to the entrance aperture of
the spectroscopic fibers), to prevent saturation and 
excessive crosstalk in the spectrographs.  Unless otherwise indicated,
the magnitudes used below are point spread function (PSF) magnitudes
for stellar and quasar 
work, and are Petrosian (1976) magnitudes for galaxy targets.
Similarly, for galaxies, colors are based on fits to exponential or de
Vaucouleurs profiles convolved with the PSF (Stoughton \etal\
2000; Abazajian \etal\ 2004; we refer to these as model colors in what
follows). All magnitudes and 
colors have been corrected for SFD Galactic extinction.  With a few
exceptions, objects for which spectroscopy already existed from
previous spectroscopic observations were excluded as targets.  This
means that defining complete samples requires combining objects from
more than one selection criterion. 

\subsection{Galactic Kinematic Programs}
\label{sec:kinematics}
  Many of the Southern plate selection algorithms were designed to
  select stars of various types, mostly for kinematic studies of the
  Galactic halo and disks.  

{\bf F stars:} \qquad F stars are numerous in the Galaxy, have sharp
spectral features allowing accurate radial velocities to be measured,
are approximate standard candles if the stars are on the main
sequence, and are of high enough luminosity that they can be seen to great
distances.  This program aims to use F
star radial velocities to understand the kinematics of the outer parts
of the Milky Way.    Most of these plates are in
the Southern Galactic Cap Equatorial Stripe (chunks {\tt fstar29} and 
{\tt fstar51}), but there are also three
plates in the vicinity of M31 ({\tt fstar72}), and another two plates centered roughly
on the Perseus cluster ({\tt seguetest84}).  The M31 (e.g., Zucker
\etal\ 2004ab) and 
Perseus imaging data are {\em not}
included in DR4, but will be included in a future data release.   A
program to target quasars in the plates near M31 is described in
\S~\ref{sec:other}; similarly, see \S~\ref{sec:lowz_gal} for a program
to target galaxies in the Perseus plates. 

For the F-star program, stellar objects with $-0.3 < (g-r) < 0.3$ and
with $19.0 < g < 20.5$ were selected.  
Given the faint
magnitudes of these stars and the desire for accurate radial
velocities, these plates were observed to a 
S/N per pixel of 5 for stars with $g = 20.2$, requiring roughly twice
the usual spectroscopic exposure of 45 minutes. 

{\bf Main Sequence Turnoff:} \qquad
This program was designed to study the kinematics and metallicities of
high-latitude thick disk and halo stars.  Stars were selected in two
SDSS spectroscopic tiles, one centered at $(l,b) = (64,-45)$ and the
other at $(l,b)=(114,-62)$.  In both tiles stars were selected to be $g-r
< 0.8$ and $r < 19.15$.  The tile at $(l,b) = 114,-62)$ is shared with
the Thick/Thin disk program described below, and there is an
additional cut at $i-z > 0.2$ to avoid overlap.  The data from this
program that appear in DR4 are a randomly-selected subset of the total
number of available targets in both tiles.  The exposure time for this
program was the standard spectroscopic exposure time for the main
survey.

{\bf Thick/Thin Disk stars:} \qquad 
A third program was focused on the kinematics of the thin and thick disks,
targeting a complete sample of bright ($i < 18.26$), red ($i-z > 0.2$) late-type
stars in three adjacent spectroscopic tiles centered at high Galactic
latitude ($l = 123^\circ$, $b = -63^\circ$).  It provides an
in situ sample of thin and thick disk stars with radial velocities, proper
motions, and spectroscopic metallicity determinations, densely sampling a
single line of sight out to 2 kpc above the Galactic plane.

{\bf SEGUE Test plates:} \qquad
SEGUE (``Sloan Extension for Galactic Understanding and
Exploration''; Newberg \etal\ 2003) is part of a follow-on project to
the SDSS, emphasizing spectra of stars to study stellar populations
and Galactic structure.  The targets on the SEGUE test plates in DR4
are Galactic stars of a variety of colors and magnitudes, meant to
sample the range of available spectral types and to test the
reproducibility of radial velocity measurements at faint magnitudes.
	SEGUE aims to probe the structure of the Milky Way
 	by sampling Horizontal Branch, F turnoff, G dwarf and K dwarf and
 	giant stars, representing a variety of distances in the
 	disk and halo.  These plates are being used to refine target
	selection for the SEGUE program itself, and will be described
	in detail in a future paper.  

\subsection{Other Stellar Target Selection Algorithms}

{\bf Spectra of Everything:} \qquad
  As part of an exploration of the full stellar locus, as well as a
  search for unusual objects of all sorts, we carried out a survey of
  all point sources.  These objects were
  assigned to available spectroscopic fibers on the so-called merged
  program plates, which included a mixture of mostly extragalactic targets.  In 2002
  (chunk 48), this included a random sampling of all point sources with
  clean photometry 
  (see the discussion of fatal and non-fatal flags by Richards \etal\
  2002) with SFD-corrected $i$-band PSF magnitudes brighter than
  19.1.  Not surprisingly, the vast majority of the targets were
  chosen from the densest core of the stellar locus in color-color
  space. 

 In chunk 73, we revised the algorithm to put greater weight on the wings of the
  stellar locus.  We defined a distance from the ridge of the stellar
  locus, by asking for the median and standard deviation $u-g$, $g-r$,
  and $i-z$ of stars in narrow bins of $r-i$.  Having tabulated these,
  we calculated a crude $\chi^2$-like quantity:
\begin{equation} 
L = {1\over 3} \sum_{u-g,g-r,i-z}
  \left({color - median\ color \over standard\ deviation}\right)^2.
\end{equation}
75\% of all stars had $L < 1$.  We sparsely sampled the objects with
$L<1$; decreasing the sampling rate for smaller $L$.  As
Table~\ref{table:southern_flags} describes, $L>1$ and $L<1$ 
objects are given different flags.  

These spectra were used for a determination of the completeness of the
quasar target selection algorithm (Vanden Berk \etal\ 2005).  The
overwhelming majority of these objects are confirmed to be stars; only 10
of the 19,543 of the Spectra of Everything targets analyzed in that
paper are quasars not previously targeted as such. 

{\bf High Proper Motion:} \qquad 
Munn \etal\ (2004) matched USNO-B astrometric data (Monet \etal\
2003) with SDSS to determine proper motions of stars
(\S~\ref{sec:new}).  We targeted for spectroscopy stars with high
proper motion along the equatorial stripe in the South Galactic Cap.
The stars selected had $r < 19.5$, and two cuts were 
made on proper motion.  The first was a simple cut of proper motion
$\mu > 100$ mas per year.  The second used a cut in reduced
proper motion $H_r$, as follows.  Define $H_r$ as:
\begin{equation} 
H_r = r + 5 \log_{10} \mu + 5.
\end{equation}
The sample was defined by:
\begin{eqnarray}
         g-i < 2.000 &\quad& H_r > 16\nonumber \\
       2.000 < g-i < 2.375 &\quad&  H_r > 8 + 4 (g-i) \\
       2.375 < g-i         &\quad&  H_r > 17.5 \nonumber
\end{eqnarray}
This cut removes the majority of main sequence red dwarfs, which
would otherwise swamp the sample.  This sample is useful for studying
the kinematics and spectral properties of very nearby objects,
especially white dwarfs and objects near the end of the main
sequence.  Despite the two cuts to define the sample, these are marked
with a single target selection flag.  

{\bf Stellar Locus:}\qquad For plates 323 and 324, stellar targets
were chosen from SDSS imaging data, randomly sampling the stellar
locus in color space, in order to explore the full range of stellar
spectra.  In particular, grids of width 0.04 magnitude in the
$(u-g),(r-i)$ and $(r-i),(i-z)$ color planes were set down, and a
single star at each grid point was chosen, if the grid had any stars
at all.  This process was repeated until each plate had 600 targets
assigned.  These objects have the target flag STAR\_BHB set.  
  
\subsection{Galaxy Target Selection Algorithms: low redshifts}
\label{sec:lowz_gal}

{\bf Main Extension: Galaxies} \qquad
In Chunk 22, observed in Fall 2001, we used direct extensions of the
main survey target selection algorithms.  In particular, the main
galaxy sample (Strauss \etal\ 2002) was modified only slightly, by
removing the cut on objects with half-light Petrosian $r$ band surface
brightness $\mu_{50,r}$ below 24.5 mag/arcsec$^2$.  This adds less than one object
per square degree. 

For LRGs, the sample was changed to probe galaxies fainter and bluer
than those of the main survey selection described by Eisenstein \etal\ (2001).
In particular, the cuts that were used garner about 40
objects/deg$^2$: 
\begin{eqnarray}
 r &<& 19.5\nonumber \\
        r &<& 13.4 + c_\parallel/0.3     \\
        \mu_{50,r} &<& 25 \nonumber \\
        |c_\perp|&<& 0.4  \nonumber
\label{eq:cut1}
\end{eqnarray}
for Cut I, and 
\begin{eqnarray}
r &<& 19.5\nonumber \\
        g-r &>& 1.65 - c_\perp\\
        \mu_{50,r} &<& 25 \nonumber \\
        r_{\rm psf}-r_{\rm model} &>& 0.3       \nonumber
\label{eq:cut2}
\end{eqnarray}
for Cut II.  Here,    $c_\parallel \equiv 0.7(g-r) + 1.2(r-i-0.18)$
and  $c_\perp \equiv (r-i) - (g-r)/8.0$.  In particular,
equation~(\ref{eq:cut1}) pushes the Cut I selection 0.3 magnitudes
fainter, while equation~(\ref{eq:cut2}) significantly loosens the
primary color cut of Cut II, to more fully explore the transition
between Cut I and Cut II.  

  Note that these plates also include an extension to the quasar
  target selection algorithm (\S~\ref{sec:other}). 

{\bf Complete Main Galaxies:} \qquad
  The main galaxy sample, as described by Strauss \etal\ (2002) is as
  complete as we can make it.  However, the subset for which we
  actually obtain a spectrum is only about 90\%.  This
  incompleteness has several causes, including the fact that two
  spectroscopic fibers cannot be placed closer than $55''$ on a given
  plate, possible gaps between the plates, fibers that fall out of
  their holes, and so on.  This program aims to observe the remaining
  10\% of galaxy targets in order to have a region of sky with truly
  complete galaxy spectroscopic coverage.  This is particularly
  important for studies of galaxy pairs, which are by definition
  strongly affected by the $55''$ rule.  The target selection
  algorithm is simple: all galaxies selected by the algorithm
  described by Strauss \etal\
  (2002) in the Southern Equatorial Stripe, minus those galaxies that
  actually have a successfully  measured redshift in routine
  targeting.  A complete sample of quasars and LRGs was similarly
  defined and targeted on the same plates. 

{\bf u-band Galaxies:} \qquad The main galaxy target selection is
carried out in the $r$ band, and is flux-limited at $r = 17.77$.  The
bluest galaxies have $u-r \approx 0.6$, thus the sample is complete
for galaxies to
$u \approx 18.4$.  In order to further explore the $u$-band luminosity
density and recent history of star formation in the universe, we
carried out a deeper $u$-band survey of galaxies in the SDSS.  The
sample consists of galaxies whose SFD-corrected magnitudes satisfy the
following criteria: $u_{\rm select}<19.8, g<20.5$, and $17.5<r<20.5$,
which had not 
been previously spectroscopically targeted.  Here, the $u$-band magnitude is
calculated from the $r$-band Petrosian magnitude and model colors:
$u_{\rm select} 
= r_{\rm Petro} + (u-r)_{\rm model}$, as this quantity has better
noise properties than does $u_{\rm Petro}$.  Like main galaxy target
selection, a cut was placed on $r$ band half-light surface brightness,
$\mu_{50,r}<24.5$, to exclude extremely low surface brightness
galaxies. Similarly, a cut on $r_{\rm psf} - r_{\rm model}>0.2$
excludes stars.  Baldry \etal\ (2005) describe the sample in detail,
and present the resulting $u$-band luminosity function.

  There were also two categories of lower-priority $u$-band selected
  targets for Chunk 73.  The ``extra'' $u$-band targets relaxed the selection
  criteria described in the previous paragraph slightly: 
\begin{eqnarray}
 u_{\rm select} < 20.0 &{\rm \ or\ }&
 u_{\rm model} < 19.8 {\rm \ or\ }  u_{\rm Petro} < 19.5, \nonumber\\
g < 20.7; & & 17.3 < r < 20.7\\
\mu_{50,r} < 24.7 & & r_{\rm psf} - r_{\rm model} > 0.15,\nonumber
\end{eqnarray}
while the ``extra2'' targets relaxed the magnitude limit further: 
\begin{eqnarray}
 u_{\rm select} &<& 20.3\nonumber\\
g < 20.5; & & 17.5 < r < 20.5,\\
\mu_{50,r} < 24.5; & & r_{\rm psf} - r_{\rm model} > 0.20.\nonumber
\end{eqnarray}

  Note that the objects targeted in this program do not include
  objects with spectroscopic observations already in hand.  Thus one
  needs to combine objects from various programs to define a complete
  sample.  See Baldry \etal\ (2005) for
  determination of the completeness of the resulting sample.

{\bf Low redshift galaxies:} \qquad
  For studies of the faint end of the galaxy luminosity function
  (e.g., Blanton \etal\ 2005b), the small-scale clustering of
  galaxies, and galaxy populations in clusters, it is useful to have a
  survey that 
  probes to as low luminosities as possible.  With the apparent
  magnitude limit of the main galaxy survey, the volume to which
  galaxies with absolute magnitudes fainter than, say, $M_r = -17$, are
  probed is quite small.  With this in mind, we carried out a survey
  of low redshift objects, selected to nearly two magnitudes fainter 
  than the SDSS main sample limit of $r = 17.77$.  Our redshift selection used
  photometric redshifts  derived from second-order polynomial
  fits to observed Petrosian $r$ magnitudes and model colors, with
  separate fits done in bins of model $g-r$ color.
  For Chunks 45, 52, and 62, we used the SDSS EDR photometry and 
  spectroscopy then available to derive photometric redshifts, 
  while for Chunks 74 and 97, we were able to derive improved
  photometric redshifts using catalog-coadded Stripe 82 SDSS 
  photometry, combined with all available SDSS redshift data on the Southern
  Equatorial Stripe as of 11 July 2003.  This included much of the
  data taken for the express purpose of calibrating the photometric
  redshift relation in the SDSS photometric system; see
  \S~\ref{sec:photoz} below.  

  Galaxies were then chosen for observations based on their
  photometric redshift $z_p$ and Petrosian magnitude  $r$.  In
  particular, the aim was to target as complete a sample as possible
  for $17.77 \leq r < 19.0$ and true redshift below 0.15, 
  and sparse samples to higher redshifts, as well as at 
  fainter magnitudes $19.0 \leq r < 19.5$.  
  The specific target categories, in order of highest to lowest priority 
  for fiber assignment, are listed in Table~\ref{table:lowz_fraction}.
  The sparse sampling fraction values were chosen to get reasonable 
  distributions of objects over the target categories and to keep 
  approximately similar target distributions from chunk to chunk, 
  which resulted in somewhat different sampling fractions for each of the
  five chunks in which this algorithm was used, as given in
  Table~\ref{table:lowz_fraction}.  
      The tabulated sampling fractions give the fraction of targets made
      available for fiber assignment; the actual fraction of targets
      with spectra is lower, in particular for the lowest priority targets.
  A caveat to note is that Chunk 45 used the star/galaxy separation
  criteria of the SDSS photometric pipeline to select galaxies, and this
  resulted in noticeable contamination of stars in several
  of the Chunk 45 plates located at lower galactic 
  latitudes.  The other chunks used the star/galaxy cut employed by
  the SDSS main sample target selection algorithm 
  ($r_{\rm PSF} - r_{\rm model} \geq 0.3$ or 0.24, depending on the version
  of the photometric pipeline), which is more conservative for
  selecting galaxies.

Two additional southern programs, both in Chunk 29,
also selected samples of low-redshift and/or low-luminosity galaxies.
One plate, 811 (chunk {\tt loveday29}), was also devoted to low-luminosity
galaxies, but 
used a different algorithm.  The sample was chosen as all objects from
the EDR photometric redshift catalog (Csabai \etal\ 2003) with $z_p >
0.003$, $r_{\rm Petro} < 20$, and estimated absolute magnitude $M_r >
-18$.  

Finally, plates 802-806 in Chunk 29 ({\tt annis29}) used a somewhat simpler
photometric  
redshift relation, and selected those objects with $i_{\rm Petro} \le
  20, i_{\rm Petro} + (r-i)_{\rm model}\ge 17.75$ and $z_p \le 0.17-0.19$.
  The photometric redshift limit on each plate was chosen to give
  enough targets to match the available number of fibers.

{\bf Perseus:} \qquad Imaging scans were taken centered roughly on the
Perseus cluster at $z \approx 0.018$, and were used to
spectroscopically target galaxies.  The double exposure plates
1665, 1666 targeted, and obtained 
redshifts for, approximately 400 galaxies as faint
as $r_{\rm fiber} = 18.8$ in a region centered on the cluster
at $(\alpha,\delta) = (49.96^\circ, 41.53^\circ)$. 
The majority of the galaxies are associated with the cluster, although
there are 50 objects in a background overdensity at $z\approx 0.05$
(Brunzendorf \& Meusinger 1999).  These plates also included
approximately 300 foreground F-stars, as described in
\S~\ref{sec:kinematics}. 

\subsection{Galaxy Target Selection Algorithms for calibrating
  Photometric Redshifts}
\label{sec:photoz}

{\bf Photo-z:}\qquad
  The SDSS five-band photometry goes substantially fainter than does the
  spectroscopy, suggesting the opportunity to derive photometric
  redshifts for vastly more objects than have spectroscopy (see, for example,
  Csabai \etal\ 2003).
  Calibrating the photometric redshift relation requires a training
  sample exploring the same range of apparent magnitudes and colors as
  the objects for which photometric redshifts will eventually be
  derived.  The SDSS LRG sample (Eisenstein \etal\ 2001) obtains
  spectra for red faint ($r < 19.5$) galaxies; photometric redshifts
  of this relatively uniform population (e.g., Eisenstein \etal\ 2003)
  are fairly robust (e.g., Padmanabhan \etal\ 2005).  However, we do
  not have a corresponding sample of faint blue galaxies for the
  calibration of photometric redshifts.  Therefore, a series of
  spectroscopic plates was designed to obtain redshifts for the blue
  end of the galaxy color distribution at the faint end.  

Galaxies were selected using the following cuts, designed to select
objects with photometric redshift greater than 0.3 (Csabai \etal\ 2003):

\begin{eqnarray}
 0.40+0.6(u-g) &<& g-r < 1.7-0.1(u-g)\nonumber \\
-0.5 < u-g < 3.0; & & 0 < g-r < 1.8 \nonumber \\
-0.5< r-i < 1.5; & & -1 < i-z < 1.5  \\
18.0 < u < 24.0; & & 18.0 < g < 21.5 \nonumber \\
17.8 < r < 19.5; & & 16.5 < i < 20.5 \nonumber \\
16.0< z < 20.0; & & \sigma_u< 0.6;\ \sigma_{g,r,i,z}<0.25.\nonumber
\end{eqnarray}

For objects that satisfied the above cuts, the quantity 
$\exp\left[c((g-r) - (0.40+0.6(u-g)))\right]$ was calculated; if it
was larger than a random number chosen between zero and one, the
object was targeted for spectroscopy.  The coefficient $c=0.1411$ was
chosen to obtain an appropriate density of targets.  Note that plates
672 and 809 have the same center, and some objects were
inadvertently observed twice.  

\subsection{Selection of higher-redshift galaxies}

{\bf Deep LRG Exposures:} \qquad 
The SDSS LRG sample (Eisenstein \etal\ 2001) targets high-redshift
($0.2 < z < 0.55$), luminous galaxies by their colors and magnitudes.
As part of Southern targeting, we have extended this algorithm in
several ways.  The first, the Deep LRG sample, uses double-length
spectroscopic exposures to get higher S/N spectra of LRGs with $z >
0.25$.  Combining with previous spectroscopy therefore gives a total
exposure time three times the standard value. 
These observations serve two purposes: first, to obtain
measurements of 
velocity dispersion for galaxies where the current, single-pass
spectroscopy is only ``good enough for a redshift''.  Second, given
the discontinuity in targeting algorithm at $z \approx 0.4$ (the
distinction between Cut I and Cut II; see Eisenstein \etal\ 2001)
higher S/N spectra allow the exploration of possible spectroscopic
changes due to differences in stellar
populations at this transition.  

{\bf Faint LRGs:} \qquad
The LRG Cut II sample aims for a flux-limited sample of LRG with
redshifts roughly between 0.40 and 0.55.  We also experimented with an
extension of this cut, going substantially fainter.  The Faint LRG
sample has the following criteria: 
\begin{eqnarray}
        17.5 < i_{\rm deV} < 20\nonumber \\
        i_{\rm Petro} < 19.1 \nonumber \\
        0.5 < g-r < 3.0      \nonumber \\
        0.0 < r-i < 2.0   \\
        c_\parallel \equiv 0.7(g-r) + 1.2(r-i-0.18) > 1.6\nonumber \\
        d_\perp \equiv (r-i) - (g-r)/8.0 > 0.5.\nonumber
\end{eqnarray}
Here, $i_{\rm deV}$ refers to the de Vaucouleurs model magnitude in the
$i$-band, and all colors are model colors.  Given the typical $r-i
\approx 0.7$ for LRGs, this is about 0.3 mag deeper than the $r <
19.5$ Cut II
sample, limited at $r = 19.5$.  There are additional cuts, to select
against, in sequence, very concentrated objects, very low surface
brightness objects, and stellar objects, respectively: 
\begin{eqnarray}
        i_{\rm fiber} - i_{\rm deV} &>& 0 \nonumber \\
        i_{\rm fiber} &<& 21.4 \nonumber \\
        i_{\rm psf} - i_{\rm model} &>& 0.15 \\
        r_{\rm psf} - r_{\rm model} &>& 0.4.    \nonumber
\end{eqnarray}

Finally, objects that were targeted as part of normal target
selection were removed.  These objects were observed on plates that received
the standard exposure times.  

{\bf BCGs:}\qquad
  A separate program explicitly targeted the brightest galaxies in
  clusters.  While LRGs are often the brightest galaxies in their
  clusters (e.g., Loh \& Strauss 2005), they need
  not be.  The so-called MaxBCG method described by Bahcall  \etal\
  (2003) searches for galaxies with
  the apparent magnitudes and colors of LRGs, together with a red
  sequence of fainter ellipticals in the vicinity (cf., Gladders \&
  Yee 2000).  The BCG program targeted BCG candidates found with this
  method, with estimated redshifts in the range $0.4 < z < 0.7$.  The
  MaxBCG algorithm was run on photometry derived from co-adding the
  detections (at the catalog level) of multiple scans of the Southern
  Equatorial Stripe.  

\subsection{Other target selection algorithms}
\label{sec:other}

{\bf Main Extension: Quasars:} \qquad
  In Chunk 22, the quasar target selection algorithm (Richards \etal\
  2002) was extended as
  follows.  For objects selected from the $ugri$ color cube, the
  magnitude limit was changed from $i = 19.1$ to 19.9, while for the
  $griz$ color cube (where high-redshift quasars are selected), we
  changed the limit from $i = 20.2$ to 20.4. 

  As Richards \etal\ (2002) describe, there are regions outside
  the stellar locus that are heavily contaminated by hot white dwarfs, M
  dwarf-white dwarf pairs (Smol\v ci\'c \etal\ 2004), and other non-quasars.  These regions are explicitly excluded
  from quasar target selection in the main survey; however, in the
  extension, they are allowed back in.  Similarly, there is a region
  of color space where $z \approx 2.7$ quasars intersect the stellar
  locus.  In the main survey, objects falling in this region are
  sparse-sampled to 10\% (to reduce the number of stars); in the
  extension, all objects falling in the mid-$z$ box defined in Richards
  \etal\ (2002) are targeted. 

{\bf Faint Quasars:}\qquad
The quasar target selection was further modified for the faint quasar
targets on the merged plates on Chunk 48.  In particular, the
following are the changes over the standard quasar target selection
algorithm described in 
Richards \etal\ (2002):
\begin{itemize} 
\item The magnitude limit is set to $i = 20.1$, rather than $i = 19.1$
  for the $ugri$ color cube; 
\item The magnitude limit for optical counterparts to FIRST (Becker
  \etal\ 1995) sources is set to $i = 20.65$, rather than $i = 19.1$;
\item The standard quasar target selection algorithm requires that the
  estimated PSF magnitude errors in $u$ and $g$ both be less than 0.1
  for UV excess sources ($u-g<0.6$).  This limit is now
  set to 0.2. 
\item For UV excess objects with $19.1 \le i < 20.1$,
  we add a requirement that $g - r < 0.7$ to minimize stellar contamination. 
\item Objects in the mid-$z$ box are excluded altogether. 
\item In addition to the usual `distance from the stellar locus'
  algorithm used to target quasars, in the main sample there are hard
  color cuts used to select high-redshift quasars in the $griz$ color
  cube.  These color cuts are not used in the Southern targeting. 
\item Finally, there were additional color cuts to reject unphysical
  objects affected by bad CCD columns; we required $g-r>-0.5,
  r-i>-0.5$, and $i-z>-0.6$. 
\end{itemize}

{\bf M31 Quasars:} \qquad
Low-redshift quasars were targeted on the M31 imaging data (Chunk 72)
using the standard quasar selection algorithm (Richards \etal\ 2002),
but excluding the 
high-redshift candidates selected from the $griz$ cube.  The confirmed
quasars can be used to probe gas in the halo of M31.  The plates used
for this program also included F star targets, as described in
\S~\ref{sec:kinematics}.

{\bf Double-Lobed Radio Sources:} \qquad
  Quasar target selection targets unresolved optical counterparts to
  FIRST radio sources, while ``serendipity'' target selection
  (Stoughton \etal\ 2002) selects FIRST sources with extended optical
  counterparts.  This works fine for compact or core-dominated radio
  sources, but is less effective for double-lobed radio sources, in
  which the optical counterpart is associated with a point between the
  two lobes.  An algorithm was developed to find these double-lobed sources,
  including allowing for the more challenging case of {\em bent}
  double sources.  In particular, pairs of FIRST sources separated by
  $90^{\prime\prime}$ or less without SDSS optical counterparts were
  identified.  Given the distance $d$ between the centers of these two
  sources, a rectangle is drawn centered at the midpoint between the
  sources, with
  dimensions $0.57d, 1.33d$, with the short axis parallel to the line
  connecting the two sources.  This box size was chosen empirically to
  include the core of a sample of bent double sources compiled by
  E. Blanton (see Blanton \etal\ 2001).  Optical counterparts with
  $r<19.8$ that fell into the box were selected as bent double
  counterparts.  A subset of those objects, which fell into a 12
  arcsec square centered on the midpoint between the sources, were
  selected as straight double counterparts.  

{\bf Variability:} \qquad
  As discussed above, the Southern Equatorial Stripe has been imaged
  multiple times in the course of the SDSS, allowing photometric
  variability to be studied.  Variable sources were selected for
  follow-up spectroscopy using pairs of observations of
unique unsaturated point sources with $i<21$, and requiring that the
changes in the $g$ and $r$ bands exceed 0.1 mag, and are at least 3 sigma
significant (using error estimates computed by the photometric
pipeline). The time difference between the two observations varied
from 56 days to 1212 days.

This target selection produced targets on the Southern
Equatorial stripe with a surface density of $\sim 10$ objects
deg$^{-2}$, after removing those objects which already had SDSS
spectra. The majority of these targets have colors consistent with
$z<2$ (UV excess) quasars and RR Lyrae stars, classifications
confirmed by the spectroscopy.  Two classes of targets were assigned,
based on apparent magnitude: the high priority objects were those with
$i$ magnitude brighter than 19.5, and the low-priority objects had
$19.5 < i < 21$.  In practice, all targets were tiled on the plates
observed.  

\begin{deluxetable}{lp{2cm}p{5cm}p{5cm}l}
\tablecaption{Special Chunks and Plates\label{table:Southern_plates}
}
\tablehead{\colhead{Chunk} & \colhead{Chunk name} &\colhead{Plates} & \colhead{Program} & \colhead{Comments}}
\startdata
    97 &lowz97& 1859 & low-z Galaxies, BCG, Deep LRG & \tablenotemark{a}\\
    83, 96 & seguetest83, seguetest96&1662-1664,1857 & SEGUE Test &\\
    84 &seguetest84&     1665-1666 & Perseus &\tablenotemark{b}\\
    79 &photoz79& 1629,1632,1633,1635 & Photo-z \\
    74 &lowz74& 1538,1539,1561-1566  & Low-z Galaxies, BCG, Deep LRG & \tablenotemark{a}\\
    73 &merged73&  1473-1476,1485-1499,1504-1506,1508-1511,1514-1518,1521-1523,1529 & Complete Main, $u$-band Galaxies, Variability,
    High Proper Motion Stars, Spectra of Everything, Fainter LRG
    &\tablenotemark{c}\\
    72 &fstar72&    1468,1471,1472 &M31: F stars, quasars&\tablenotemark{b} \\
    62 &lowz62&  1243 &Low-z galaxies, Deep LRG & \tablenotemark{a} \\
    52 &lowz52&  1156,1157 &Low-z galaxies, Deep LRG & \tablenotemark{a} \\
    51 &fstar51&  1149,1150,1152--1155 & F stars & \tablenotemark{a} \\
    50 &rockosi50&  1133-1135,1137,1143 & MS Turnoff stars \\
    49 &munn49&  1118-1132 & Thick/Thin Disk \\
    48 &merged48&  1062-1096,1101,1103-1107,1114-1117 & Complete Main,
    $u$-band Galaxies, Double-Lobed 
    Radio Sources, Faint Quasars, Spectra of Everything
    &\tablenotemark{c}\\
    45 &lowz45&  1021-1024,1026-1037 &Low-z galaxies, Deep LRG & \tablenotemark{a} \\
    29 &photoz29&    807-810 & Photo-z \\
    29 &annis29&     802-806 & Low-z Galaxies \\
    29 &loveday29&     811 & Low-z Galaxies \\
    29 &fstar29& 797 & F stars & \\
    22 &south22&     673-714 & Main Extension: Galaxies,Quasars&
    \tablenotemark{d}\\ 
    21 &photoz21&    669-672 & Photo-z \\
    ---&Schlegel/Locus& 323-324 & Stellar Locus & \\
\enddata

\tablenotetext{a}{Double-length exposures}
\tablenotetext{b}{Imaging not available}
\tablenotetext{c}{Standard merged program}
\tablenotetext{d}{Extensions of standard algorithms}
\end{deluxetable}

\begin{deluxetable}{lllcr}
\tabletypesize{\footnotesize}
\tablecaption{Target Selection Flags\label{table:southern_flags}}
\tablehead{\colhead{Chunk} &
\colhead{Target} &
\colhead{primTarget}& \colhead{Number} &\colhead{Comments} }
\startdata

84 &
Perseus: Galaxies &
0x80000040 
& 787 & \tablenotemark{a}\\

84 &
Perseus: F stars &
0x80002000 
& 313 & \tablenotemark{a}\\

83,96 &
SEGUE Test &\nodata
& 2116  &\tablenotemark{a,b} \\

74,97 &
BCG &
0x80000100
& 439  &\tablenotemark{a}\\

73 &
Complete Main &
\nodata
& 2020  &\tablenotemark{d} \\

73 &
u-band Selected Galaxies: priority &
0x80000040
& 765  &\tablenotemark{a}\\

73 &
u-band Selected Galaxies: extra &
0x800000C0
& 2128  &\tablenotemark{a}\\

73 &
u-band Selected Galaxies: extra2 &
0x80000140
& 2108  &\tablenotemark{a}\\

73 &
Variability: high priority &
0x81000000
& 318  &\tablenotemark{a}\\

73 &
Variability: low priority &
0x80800000
& 660  &\tablenotemark{a}\\

73 &
High Proper Motion Stars &
0x80010000
& 966  &\tablenotemark{a}\\

73 &
Spectra of Everything: $L> 1$ &
0x80000001
& 3934  &\tablenotemark{a}\\

73 &
Spectra of Everything: $L< 1$ &
0x80000002
& 5478  &\tablenotemark{a}\\

73 &
Faint LRG &
0x840000A0
& 2301  &\tablenotemark{a}\\

72 &
M31: F stars &
0x80002000
& 1623  &\tablenotemark{a}\\

72 &
M31: Quasars &
0x80000004
& 144  &\tablenotemark{a}\\

51 &
F Star &
0x80002000
& 3531  &\tablenotemark{a}\\

50 &
MS Turnoff &
0x80002000
& 2947  &\tablenotemark{a}\\

49 &
Thick/Thin Disk &
0x80040000
& 8859  &\tablenotemark{a}\\

48 &
Complete Main &
\nodata
& 1605  &\tablenotemark{d} \\

48 &
u-band Selected Galaxies &
0x80000040
& 3426  &\tablenotemark{a}\\

48 &
Double Radio Sources/Bent &
0x80200000
& 97  &\tablenotemark{a}\\

48 &
Double Radio Sources/Straight &
0x80000010
& 13  &\tablenotemark{a}\\

48 &
Faint Southern Quasars &
\nodata 
& 6655 & \tablenotemark{a,b}\\

48 &
Spectra of Everything &
0xA0000000
& 14779  &\tablenotemark{a}\\

45,52,62,74,97 &
Low-z &
0x80000040
& 14210  &\tablenotemark{a}\\

45,52,62,74,97 &
Deep LRG/Cut I &
0x80000020
& 1230  &\tablenotemark{a}\\

45,52,62,74,97 &
Deep LRG/Cut II &
0x84000020
& 205  &\tablenotemark{a}\\

29 &
Low-z (Loveday) &
0x80000040
& 570  &\tablenotemark{a}\\

29 &
Low-z (Annis) &
0x80000040
& 2849  &\tablenotemark{a}\\

29 &
F Star &
0
& 583  &\tablenotemark{e}\\

22 &
Main & \nodata
& 8008  &\tablenotemark{d}\\

22 &
Main Extension & \nodata
& 16760  &\tablenotemark{b,c}\\

21,29,79 &
Photo-z &
0x80000040 
& 4526 & \tablenotemark{a}\\

\nodata & Stellar Locus & 
0x00002000
& 1188  &\tablenotemark{e}\\

\enddata

\tablenotetext{a}{The Southern plates are flagged with the highest bit
set in secTarget; i.e., secTarget = 0x80000000}
\tablenotetext{b}{Regular primTarget flags, with the highest bit set}
\tablenotetext{c}{Regular secTarget flags, with the highest bit set}
\tablenotetext{d}{Regular primTarget and secTarget flags}
\tablenotetext{e}{Stars of interest are flagged STAR\_BHB in the
  Objtype\_name column of specObjAll in the CAS.}

\end{deluxetable}

\section{Other New Features of DR4}
\label{sec:new}

The DR3 paper (Abazajian \etal\ 2005) described a quality flag
calculated for each field in the SDSS imaging data.  This flag was
based on the measured point spread function (PSF), the difference between
aperture and PSF magnitudes of stars (a diagnostic of errors in the
determination of the PSF), and the location of the stellar locus in
color-color space\footnote{Abazajian \etal\ (2005) described a bug
  whereby the quality was based on only the $s$ color (Helmi \etal\
  2003), as opposed to all four available principal colors which
  describe the stellar locus.  Unfortunately, this bug has not yet
  been fixed as of this writing.}.  With DR4, we make available the
detailed plots on 
which these quantities are based.  Much of the rationale of the
quantities plotted is given by Ivezi\'c \etal\ (2004b).  For each of
the 204 imaging runs included in DR4, a web page is available with
detailed statistics and plots of the quantities used in the
determination of the quality, as well as tests of the uniformity of
the sky levels from one camera column to another, and statistics on
the relative astrometry between the different photometric bands.
These statistics are summarized in a table giving the fraction of the
fields in each run of a given quality; these then allow an overall
determination of the quality of each run.  85\% of the 204 runs show
median offsets in the $s$ principal color between runs of less than 0.02
magnitudes in all six camera columns. 

The DR2 paper (Abazajian \etal\ 2004) describes the availability of
proper motions for each object based on matching with USNO-B (Monet
\etal\ 2003). The DR4 CAS also includes the recalibrated proper
motions based on using the SDSS astrometry to improve the USNO-B
positions (Munn \etal\ 2004).  

\section{Object Counts Near Bright Objects}
\label{sec:caveats}

A study of weak lensing in the SDSS imaging data (Mandelbaum \etal\
2005) found a systematic, 5\% decrease in the number density of faint
objects within $90''$ of bright ($r < 18$) galaxies.  This was found
to be due to a systematic overestimation of the sky levels in the
vicinity of bright objects, which of course will affect all the
measured photometric quantities of faint objects, including the
classification as stars and galaxies.  The brighter the object, the
larger the overestimation of sky; indeed, restricting to foreground
galaxies brighter than $r = 16$, the number density of fainter
galaxies is 10\% below the mean.  This is due to the way in
which the sky levels are estimated in the SDSS; they can be biased
upward by the faint outer isophotes of a bright galaxy (Stoughton
\etal\ 2002; see also the discussion by Strauss \etal\ 2002).  Closer
than $40''$ to the bright objects, the intrinsic clustering of
galaxies makes it difficult to assess the problem.  This problem is
most relevant for applications that involve correlating positions of
bright objects with faint ones, yielding a spurious anti-correlation on
the affected scales.  We are currently investigating changes to the
imaging pipeline to address this problem.

\section{Conclusions}
\label{sec:conclusions}
  We have presented the SDSS Fourth Data Release, consisting of
  five-band imaging data over 6670 deg$^2$ and spectra for over
  800,000 objects.  These data represent roughly a 25\% increment over
  the previous data release (DR3, Abazajian \etal\ 2005).  

  The next data release is planned for mid-2006, and will consist of
  all SDSS survey-quality data gathered through June 2005.  The
  immediate goals of the SDSS are to fill the gap between the two
  portions of the Northern Galactic Hemisphere seen in
  Figure~\ref{fig:skydist} in both imaging and spectroscopy.  This,
  together with new surveys to study structure in the Milky Way
  (Newberg \etal\ 2003) and supernovae at $0.1 < z < 0.3$ (Frieman
  \etal\ 2003), will keep
  the SDSS facilities busy through Summer 2008. 

Funding for the creation and distribution of the SDSS Archive has been
provided by the Alfred P. Sloan Foundation, the Participating
Institutions, the National Aeronautics and Space Administration, the
National Science Foundation, the U.S. Department of Energy, the
Japanese Monbukagakusho, and the Max Planck Society.  The SDSS Web
site is http://www.sdss.org/.

The SDSS is managed by the Astrophysical Research Consortium (ARC) for
the Participating Institutions.  The Participating Institutions are
The University of Chicago, Fermilab, the Institute for Advanced Study,
the Japan Participation Group, The Johns Hopkins University,  the
Korean Scientist Group, Los Alamos National Laboratory, the
Max-Planck-Institute for Astronomy (MPIA), the Max-Planck-Institute
for Astrophysics (MPA), New Mexico State University, University of
Pittsburgh,  University of Portsmouth,  Princeton University, the
United States Naval Observatory, and the University of Washington.

\end{document}